\documentclass[12pt,epsfig]{article}
\usepackage{amsmath}
\usepackage{amssymb}
\usepackage{graphicx}
\usepackage{epsfig}
\newcommand{\bea}{\begin{eqnarray}}
\newcommand{\nn}{\nonumber}
\newcommand{\eea}{\end{eqnarray}}
\newcommand{\pcl}{\pi_{cl}}
\newcommand{\ep}{\epsilon}
\newcommand{\omc}{\omega_{cl}}

\def\bes{\begin{eqnarray}}
 \def\ees{\end{eqnarray}}
\def\be{\begin{equation}}
\def\ee{\end{equation}}
\def\bs{\begin{subequations}}
\def\es{\end{subequations}}
\newcommand{\een}{\end{subequations}}
\newcommand{\ben}{\begin{subequations}}
\newcommand{\beq}{\begin{eqalignno}}
\newcommand{\eeq}{\end{eqalignno}}

\def\trr{{\rm tr}}

 \def\picl{{\pi_{cl}}}
 \def\dpi{{\delta\pi}}
 \def\dG{{\delta G}}

 \def\ex{\epsilon}
 
 \def\Lx{\Lambda}

 \def\Tc{{\mathcal T}}

\newcommand{\calk}{{\cal K}}

\def \lta {\mathrel{\vcenter
     {\hbox{$<$}\nointerlineskip\hbox{$\sim$}}}}

\begin{document}

\flushright{CERN-PH-TH/2013-303}

\begin{center}
{ \Large \bf
Suppression of Quantum Corrections\\  by Classical Backgrounds}
\\
\vspace{1.5cm}
{\Large 
Nikolaos Brouzakis$^1$  and Nikolaos Tetradis$^{1,2}$ 
} 
\\
\vspace{0.5cm}
{\it
$^1$ Department of Physics, University of Athens,
Zographou 157 84, Greece
\\
$^2$ Department of Physics, CERN - Theory Division, CH-1211 Geneva 23, Switzerland
} 
\end{center}
\vspace{3cm}
\abstract{
We use heat-kernel techniques in order to compute the one-loop effective action in the cubic 
Galileon theory for a background that realizes the Vainshtein mechanism. We find that the UV divergences
are suppressed relative to the predictions of standard perturbation theory at length scales below the Vainshtein radius. 
}
\newpage

\section{Introduction} \label{intro}

Higher--derivative theories are perturbatively nonrenormalizable. As a result, their predictivity is limited by the necessity 
to introduce an infinite number of counterterms in order to cancel the ultraviolet (UV) divergences appearing in the quantum corrections.
Such theories can still be treated as effective below an energy scale $\Lx$ suppressing the couplings in the nonrenormalizable terms. 
If the UV completion of the theory at the scale $\Lx$ is not known, one must include all the effective terms allowed by the low-energy symmetries. 
Despite these general expectations, it is still possible that the UV behavior of the theory may be improved 
through a rearrangement of the perturbative  expansion, or at the nonperturbative level.
For example, one could incorporate some of the higher-derivative terms in an effective propagator. In 
Fourier space the propagator would then fall much faster than the standard one for increasing momenta, so that the UV divergences could 
be reduced or eliminated. However, this approach does not have internal consistency \cite{weinberg}.
The additional terms incorporated in the propagator become relevant near the UV scale $\Lx$. It is impossible to justify the exclusion of 
terms with even more derivatives, which could give larger contributions near $\Lx$. 

We are interested in a different aspect of the quantum theory: The possibility that the classical background
around which the fields are expanded can reduce the magnitude of quantum corrections. This scenario makes sense only for
inhomogeneous backgrounds, as in the opposite case the effect amounts to a simple redefinition of scales. A specific example we
have in mind involves the cubic Galileon theory, which describes the dynamics of the scalar mode that 
survives in the decoupling limit of the DGP model \cite{dgp}. The action contains a higher-derivative
term, cubic in the field $\pi(x)$, with a dimensionful coupling that sets the scale $\Lx$ at which the theory becomes strongly
coupled.
The tree-level action in Euclidean space is  
\be
S_0=\int d^4x \left\{ \frac{1}{2} (\partial \pi)^2 -\frac{\nu}{2} (\partial \pi)^2 \Box \pi \right\},
\label{cubic} \ee 
with $\nu\sim 1/\Lx^3$. 
 The action is invariant under the Galilean transformation
$\pi(x)\to \pi(x)+b_\mu x^\mu+c$, up to surface terms.  
Despite the presence of four derivatives in the second term, the equation of motion is a second-order partial
differential equation. This property, which guarantees the absence of ghosts in the spectrum in the trivial vacuum, is also preserved 
within the Galileon theory, which includes a finite number of higher-order terms \cite{galileon}.

The Galileon theory can provide a realization of the Vainshtein mechanism, which
has been introduced in order to suppress the propagation of the physical mode of the massive graviton that survives in the limit of 
vanishing mass \cite{vainshtein}. The cubic theory of eq. (\ref{cubic}) has a spherically symmetric solution 
$\pi_{cl}=\picl(w)$, with $w=r^2$, given by
\be
\pi'_{cl}(w)=\frac{1}{8\nu}\left(1-\sqrt{1+ \frac{16\nu c}{w^{3/2}}} \right),
\label{solgsta} \ee
where the prime denotes a derivative with respect to $w$ and we have assumed that $c,\nu>0$.
For $w\gg w_V$, with $w_V=r^2_V\sim (\nu c)^{2/3}$ the square of the Vainshtein radius, 
the solution is $\pi'_{cl}\sim c \,w^{-3/2}$,  so that 
$\pi_{cl}\sim c/r$. On the other hand, for $w\ll w_V$, we have $\pi'_{cl}\sim \sqrt{c/\nu}\, w^{-3/4}$,  so that 
$\pi_{cl}\sim \sqrt{c/\nu} \,\sqrt{r}$. This solution requires the presence of a large
point-like source at the origin, with strength depending on $c$. The classical fluctuations $\dpi$ of the field around a general background $\pi_{cl}$ 
obey the linearized equation $\Delta_{cl} \dpi=0$, with the operator
\be
\Delta=-\Box+2\nu \left( \Box \pi \right) \Box -2\nu \left( \partial_{\mu}\partial_{\nu} \pi\right) \partial^{\mu}\partial^{\nu}
\label{lapll} \ee 
evaluated for $\pi=\pi_{cl}$.
(We employ covariant notation, even though we work in Euclidean space.) 
For the background (\ref{solgsta}), the first term dominates at distances much larger than the Vainshtein radius, so that the
fluctuations $\dpi$ propagate as free waves. On the other hand, 
the dominance of the last two terms  at distances smaller than the Vainshtein
radius, where $\nu\Box \picl \gg 1$, results in the suppression of the classical fluctuations.

In ref. \cite{quantum1} it was argued that the same mechanism can lead to the suppression of quantum fluctuations as well, thus
reducing the effect of quantum corrections at the scales at which the Vainshtein mechanism operates. The essence of the argument
is that the higher-derivative terms generate a large effective wavefunction renormalization $Z$ for the fluctuation $\dpi$. If this can be absorbed
in the definition of a canonically normalized field, the couplings of the theory are reduced by powers of $Z$. Even though this intuitive 
argument seems reasonable, it is not rigorous because of the position dependence of $Z$ in the background (\ref{solgsta}). 
In this work we introduce an appropriate modification of the heat-kernel calculation of the one-loop corrections in order to 
examine the issue through a more rigorous approach. 

In section \ref{perturb} we show how the known perturbative results for the cubic Galileon theory are reproduced through the heat kernel. 
In section \ref{back} we introduce a modificiation of the heat-kernel calculation that accounts for the effect of the background more
efficiently than perturbation theory. In this way we demontrate that the background can suppress the quantum corrections in the
region where the Vainshtein mechanism operates. In section \ref{general} we consider the general structure of the quantum corrections 
and their suppression by the background. Finally, in section \ref{discussion} we present our conclusions.

\section{Perturbation theory} \label{perturb}

Our task is to evaluate the one-loop effective action 
\be
\Gamma_{1}= \frac{1}{2} {\rm tr} \log \Delta,
\label{gamma1} \ee
where the operator $\Delta$ is given by eq. (\ref{lapll}).
Following ref. \cite{nepomechie}, we calculate the heat kernel of $\Delta$ through the relation
\be 
h(x,x',\ep)= \int \frac{d^4k}{(2 \pi)^4} e^{-ikx'}e^{-\ep \Delta} e^{ikx}.
\label{heatk} \ee
The effective action can be obtained from the diagonal part of the heat kernel as
\be
\Gamma_1=-\frac{1}{2}\int_{1/\Lx^2}^{\infty} \frac{d \ex}{\ex} \int d^4x\,  h(x,x,\ex).
\label{aaa} \ee
A lower limit has been introduced for the $\ex$-integration in order to regulate the possible UV divergences.

The higher-derivative terms in the effective action are generated through the expansion of the exponential in eq. (\ref{heatk}).
The operators act either on functions, such as $\Box \pi$, appearing in $\Delta$, or on $\exp(ikx)$.
An efficient way of carrying out the expansion is implied by the analysis of ref. \cite{nepomechie}. The integrant of 
eq. (\ref{heatk}) can be viewed as an operator acting on an arbitrary function $f(x)$. After the expansion of the exponential is performed, one
sets $f(x)=1$ in order to retain only the terms that are relevant for the evaluation of the heat kernel.
In this process we employ the operator  identities
\be 
e^{-ikx} (-\Box)e^{ikx}=k^2-2ik^\mu \partial_\mu -\Box,
\label{id2} \ee
\be
e^{-ikx} \partial_\mu \partial_\nu e^{ikx}= -k_\mu k_\nu+ik_\mu\partial_\nu+ik_\nu\partial_\mu+\partial_\mu\partial_\nu.
\label{id3} \ee
In order to determine the UV divergences, which appear for $\ex\to 0$, 
it is useful to rescale $k$ in eq. (\ref{exponent}) by $\sqrt{\ex}$, as was done in ref. \cite{nepomechie}. 
The diagonal part of the heat kernel becomes
\begin{eqnarray}
h(x,x,\ep)&=&\int\frac{d^4k}{(2 \pi)^4}\frac{1}{\ex^2}\exp \biggl\{
 - k^2+2i\sqrt{\ex} k^\mu\partial_\mu+\ex\Box+2\nu\Box\pi\left(k^2-2i\sqrt{\ex}k^\mu\partial_\mu-\ex \Box\right)
\nonumber \\
&&
~~~~~~~~~~~~~~~~~~~~~~~~~~~~~~~~~
-2\nu \partial_\mu\partial_\nu \pi \left(k^\mu k^\nu-2i\sqrt{\ex}k^\mu\partial^\nu-\ex \partial^\mu\partial^\nu  \right) 
\Bigr] \biggr\},
\label{diagon}
\end{eqnarray}
with the implicit assumption that it will be evaluated through its action on $f(x)=1$.

The standard procedure is to isolate the term $\exp(-k^2)$ and expand the rest of the exponential.
The $k$-integration can be performed with the help of formulae such as
\be
\int \frac{d^nk}{(2\pi)^n}e^{- k^2}k^\mu k^\nu k^\rho k^\sigma =\frac{1}{(4\ex\pi)^{n/2}}\frac{1}{4}
\left( g^{\mu\nu} g^{\rho\sigma}+g^{\mu\rho} g^{\nu\sigma}+g^{\mu\sigma} g^{\nu\rho} \right).
\label{iden2} \ee
The results of perturbation theory are obtained through a double expansion in $\nu$ and $\ex$.
For a given power of $\nu$, the lower-order terms in $\ex$, up to $\ex^2$, reproduce the UV divergences of the effective action. 
For example, at order $\nu^2$ the leading divergence is associated with a term $\sim (\Box \pi)^2$. 
The corresponding diagonal part of the heat kernel is
\be
h(x,x,\ex)=\frac{15}{32\pi^2\ex^2}\nu^2  (\Box \pi)^2
\label{heat1} \ee
and the contribution to the effective action 
\be
\Gamma_1^{(2)}=-\frac{1}{2}\int_{1/\Lx^2}^{\infty} \frac{d \ex}{\ex} \int d^4x\,  h(x,x,\ex)=-
\frac{15}{128\pi^2}\nu^2 \Lx^4 \int d^4x \, (\Box \pi)^2. 
\label{eff1} \ee

The heat-kernel analysis is consistent with the expectations for the quantum corrections in the Galileon theory. The structure of the 
divergent terms in the one-loop effective action is, schematically, \cite{quantum1,nonrenorm,multifield}
\be
\Gamma_1 \sim \int d^4x \sum_m \left[ 
\Lx^4+\Lx^2\partial^2+\partial^4 \log \left(\frac{\partial^2}{\Lx^2} \right)
\right] \left( \nu \partial^2 \pi \right)^m.
\label{schem} \ee 
The result (\ref{eff1}) reproduces
the leading divergence in eq. (\ref{schem}) for $m=2$. 
An explicit calculation, 
carried out in ref. \cite{quantum2} for the theory of eq. (\ref{cubic}) through dimensional regularization, reproduced the logarithmic term. 
A similar calculation of all the terms with $m=2$ was performed in ref. \cite{ctzb}. 
The problem with the effective action (\ref{schem}) is that it cannot be trusted in the region below the Vainshtein radius, where
$\nu \Box \picl \gg 1$. 
If we expand the field as $\pi=\pi_{cl}(x)+\dpi(x)$ in eq. (\ref{schem}), with the perturbation $\dpi$ assumed to be small, 
$\nu \Box \picl$ would act as an effective expansion parameter. For example, a series of interaction
terms $\sim \nu^2 \Lambda^4 (\nu \Box \picl)^{n} (\Box \dpi)^{2}$
would be generated. For the series in $n$ to converge, $\nu \Box \picl$ 
should be smaller than 1. In the opposite case, the UV divergences of the theory seem to be enhanced by the presence of the background.
In order to overcome this problem, 
we shall reformulate the calculation of the heat kernel in a way that accounts more efficiently for the influence of the background
for large values of $\nu\Box \picl$.

\section{The effect of the background} \label{back}

In order to investigate the effect of the background on the UV divergences, 
we split the field  in eq. (\ref{diagon}) as $\pi=\pcl+\dpi$ and consider the correlation functions of $\dpi$.
We  define a ``metric" 
\be 
G_{\mu \nu}=g_{\mu \nu}-2\nu\Box \pcl\, g_{\mu \nu}+2 \nu \partial_\mu \partial_\nu \pcl
\label{metric}
\ee
and the operators
\bea
D_\ex(k)&=&-2i\sqrt{\ex}k^\mu \partial_{\mu}-\ex \Box 
\label{op1}\\
L_\ex^{\mu \nu}(k)&=&2i\sqrt{\ex}k^\mu\partial^\nu+\ex \partial^\mu\partial^\nu.
\label{op2}  \eea
The exponent in  eq. (\ref{diagon}) becomes 
\begin{eqnarray}
 F&=&-G_{\mu \nu}k^{\mu}k^\nu
-(1-2\nu \Box \pcl) D_\ex(k)
+ 2 \nu \partial_\mu \partial_\nu \pcl \, L_\ex^{\mu \nu}(k)
 \nn \\   
&&~~~~~~~~~~~~
+2\nu\Box \dpi \left(k^2+D_\ex(k)\right)  
+2 \nu \partial_\mu \partial_\nu \dpi  \left(-k^{\mu} k^{\nu}+L_\ex^{\mu \nu}(k)   \right).
\label{exponent} \end{eqnarray}

We have seen that the expansion of the heat kernel in  powers of $\ex$ reproduces the UV divergences of the theory.
The combination $\nu \Box \picl$ in eq. (\ref{exponent})
is the classical expansion parameter, visible also in eq. (\ref{schem}). 
This parameter becomes large below the Vainshtein radius \cite{quantum1}, which implies that it does not generate a convergent series. 
On the other hand, $\dpi$ can be viewed as a second expansion parameter, apart from $\ex$, with the 
term $\nu \Box \dpi$ assumed to be small.
Within this scheme, the ``metric" $G_{\mu\nu}$ includes the terms of zeroth order both 
in $\sqrt{\ex}$ and $\dpi$. All such terms must be treated on equal footing, and this is accomplished by our way of
evaluating the heat kernel. The main technical difficulty is that the momentum integration in eq. (\ref{diagon}) 
cannot be performed easily for general $G_{\mu\nu}$. 
However, we can render 
the ``metric" $G_{\mu \nu}$ trivial in eq. (\ref{exponent}) by rescaling the momenta as $k^\mu=S^\mu_{~\nu} k'^\nu$,
with $S^\mu_{~\nu}$ satisfying
\be
S^\mu_{~\rho} G_{\mu\nu}S^\nu_{~\sigma}=g_{\rho\sigma}.
\label{diag} \ee
Through differentiation of this relation, $x$-derivatives of $G_{\mu\nu}$ can be expressed in terms of derivatives of $S^\mu_{~\nu}$. 
Moreover, for a trivial $g=\mathbb{I}$, we have that ${\Tc}\equiv S^TS=G^{-1}$. 
The first term of eq. (\ref{exponent}) now takes the simple form $-k'^2$. It is not possible, however, 
to isolate a term $\exp(-k'^2)$ in the heat kernel and expand the rest of the exponential. 
The reason is that $k'^\mu$ does not commute with the
derivative operators in $F$, because it contains the function $S^\mu_{~\nu}(x)$. The
Baker-Campbell-Hausdorff formula 
\be
e^{X+Y}=e^X e^Y e^{-\frac{1}{2}[X,Y]} e^{\frac{1}{6}\left(2[Y,[X,Y]]+[X,[X,Y]] \right)} ...
\label{bch} \ee
must be employed, with $X=-G_{\mu \nu}k^{\mu}k^\nu$ and $Y$ consisting of the remaining terms in eq. (\ref{exponent}).
Then, each of the exponentials, apart from the first one, must be expanded, the momenta $k$ rescaled and 
the $k'$-integrations carried out.

\begin{figure}[t]
\begin{center}
 \includegraphics[width=15.0cm]{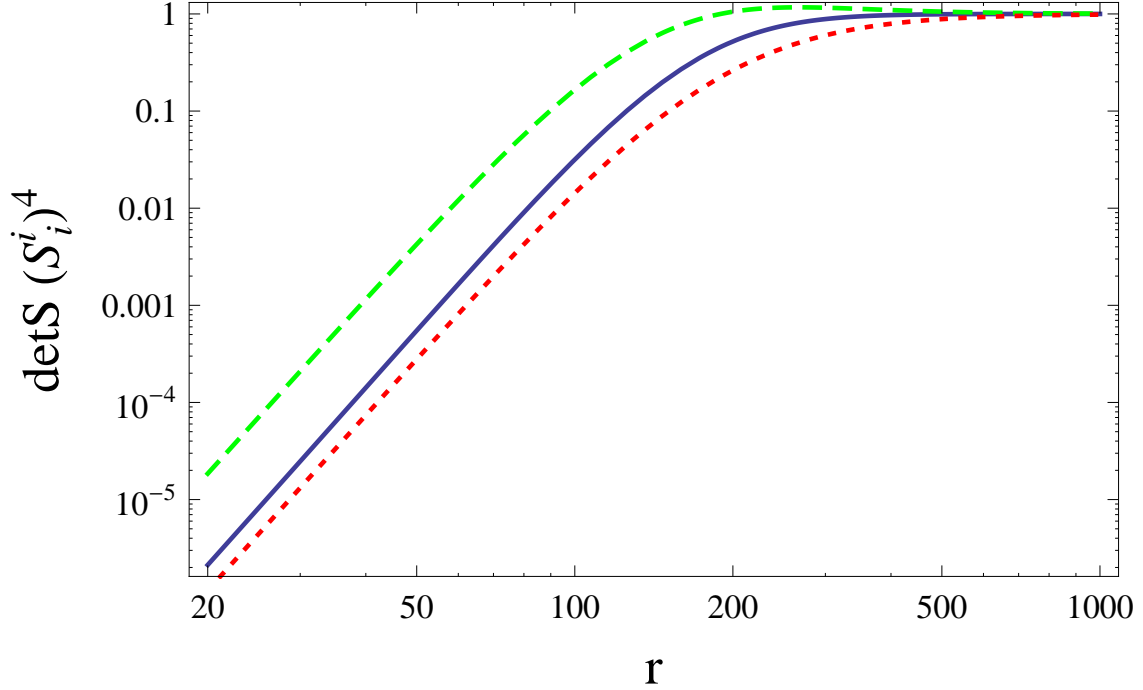}%
\end{center}
\caption{ \it $(\det S)  \left(S^i_{\, i}\right)^4 $ 
as a function of $r$ for the background of eq. (\ref{solgsta}) with $\nu=1$, $c=10^6$. 
The solid, blue line corresponds to $i=0$, the dotted, red line to $i=1$ and 
the  dashed, green line to $i=2$ or $3$. }
 \label{fig1}
 \end{figure}

We focus first on the leading divergence in the effective action, which can be obtained by observing that the $\ex$-independent terms 
in eq. (\ref{exponent}) do not include derivative operators and commute with $-G_{\mu \nu}k^{\mu}k^\nu$.
As a result, the contribution to the diagonal part of the heat kernel which is quadratic in $\dpi$ and contains the quartic divergence is 
\be 
 h(x,x,\ep)= \int \frac{d^4k}{ (2\pi)^4 }  (\det S)\frac{1}{2\ex^2} e^{-  k^2}
\left(2\nu \Box \dpi (Sk)^2+2\nu\partial_\mu \partial_\nu \dpi \left(-Sk^\mu Sk^\nu \right) \right)^2,
\label{heat3} \ee 
where we have dropped the prime on $k$.
For a spherically symmetric background 
$\picl=f(r^2)$, we consider a Cartesian system of coordinates with one of its axes along the radial direction.
We obtain
\be
G_{\mu \nu}={\rm diag} \left[
1-\nu\left( 12f'+8 r^2 f'' \right), 
1-8\nu f', 
1-\nu\left( 8f'+8 r^2 f'' \right), 
1-\nu\left( 8f'+8 r^2 f'' \right)
\right],
\label{cart} \ee
where the first entry corresponds to the time component, the second to the radial, and the last two to the 
components perpendicular to the radial.
We easily find that
\begin{eqnarray}
S^\mu_{~\nu}&=&{\rm diag} \Bigl[
\left( 1-\nu \left(  12f'+8 r^2 f'' \right) \right)^{-1/2}, 
\left( 1-8\nu f' \right)^{-1/2},
\nonumber \\ 
&&~~~~~~~~~~~~~~~~~~~~~~~~~~~\left( 1-\nu\left( 8f'+8 r^2 f'' \right) \right)^{-1/2}, 
\left( 1-\nu\left( 8f'+8 r^2 f'' \right) \right)^{-1/2}
\Bigr].
\label{cart2} \end{eqnarray}
The Jacobian determinant of the transformation is $\det S$.
After performing the  momentum integration, the contribution to the diagonal part of the heat kernel can be put in the form 
\be
h(x,x,\ep)= \frac{1}{32 \pi^2\ep^2}\nu^2 \left( \left(\Box \dpi \right)^2 P(r^2) 
- 2 (\Box \dpi)(\partial_\mu  \partial_\nu \dpi) \, V^{\mu \nu}(r^2)
+ \left( \partial_\mu  \partial_\nu \dpi\right) \left( \partial_\rho  \partial_\sigma \dpi \right)\, W^{\mu \nu \rho \sigma}(r^2)  \right),            
\label{heat4} \ee
where 
\bea
P(r^2)&=& (\det S) \left[ \left( \trr\left( {\Tc} \right)\right)^2+2\trr\left(\Tc^2 \right) \right]
\label{heat41} \\
V^{\mu \nu}(r^2)&=&  (\det S)  \left[ \trr\left(\Tc\right)\Tc^{\mu \nu}+
2\left(\Tc^2 \right)^{\mu \nu} \right]
\label{heat42} \\
W^{\mu \nu \rho \sigma}(r^2)&=& (\det S)  \left[ \Tc^{\mu \nu}\Tc^{\rho \sigma}+
2\Tc^{\mu \rho}\Tc^{\nu \sigma} \right],
\label{heat43} \eea 
with $\Tc\equiv S^TS=G^{-1}$.
Using eq. (\ref{aaa}), we find the contribution to the effective action 
\bea
\Gamma_1^{(2)} &=&
-\frac{1}{128\pi^2}\nu^2 \Lx^4 \int d^4x \,  
\Bigl( \left(\Box \dpi \right)^2 P(r^2) 
\nonumber \\
&&~~~~~~~~~~~~~~~
-2 ( \Box \dpi)( \partial_\mu  \partial_\nu \dpi) \, V^{\mu \nu}(r^2)
+ \left( \partial_\mu  \partial_\nu \dpi\right) \left( \partial_\rho  \partial_\sigma \dpi\right)\, W^{\mu \nu \rho \sigma}(r^2)  \Bigr).
\label{eff2} \eea

It is apparent from eq. (\ref{eff2}) that the invariance under the Euclidean group is broken by the background.
For a homogeneous background, for which $S$ is the four-dimensional unit matrix, eq. (\ref{eff2}) reproduces eq. (\ref{eff1}).
On the other hand, if the effective action is evaluated around the background of eq. (\ref{solgsta}), the effective Lagrangian density
has a very strong radial dependence. In order to obtain a pictorial representation of the $r$-dependence, we observe that the 
functions
$P$, $V^{\mu\nu}$, $W^{\mu\nu\rho\sigma}$ involve fourth powers of the matrix $S$ and are also proportional to its determinant.
In fig. (\ref{fig1}) we display the product of the determinant of $S$ and the fourth power of each of its diagonal elements for 
a background given by eq. (\ref{solgsta}) with $\nu=1$ and $c=10^6$. All dimensionful quantities are measured in units of the
fundamental scale $\Lx$. The Vainshtein radius is $r_V\sim (\nu c)^{1/3}=100$. 
It is apparent that the quantum corrections are suppressed below $r_V$. We estimate that $(\det S) \left( S^i_{~i}\right)^4$,
with $i=0,1,2$ or 3,
scales as $r^6/(\nu c)^2\sim (r/r_V)^6$, a behavior that is verified by fig. (\ref{fig1}). 
A substantial suppression, by several orders of magnitude, is expected for 
$1/\Lx  \lta r \lta  r_V  $.

Apart from the term we considered, there is an infinite number of higher-derivative terms, quadratic in $\dpi$,
with possible UV divergences. These result
from the expansion of the exponential in eq. (\ref{diagon}) in the way we described in the beginning of this section.
Certain features are apparent:
\begin{itemize}
\item
The momentum integration factor in the heat kernel generates a factor $\ex^{-2}$ after the rescaling, while the relation to the effective action 
involves the integration factor $d\ex/\ex$.
This means that the divergences in the effective action result from terms in the expansion of the exponential of
(\ref{exponent}) with powers of $\ex$ up to 2.
\item
 The suppression of the quantum corrections by the background arises through the matrix $S$ that rescales the momenta. 
In the region where the Vainshtein mechanism operates, the elements of $S$ have typical values 
$\sim |\nu \Box \picl|^{-1/2}\sim (r/r_V)^{3/4} \ll 1$. 
The Lagrangian density is also multiplied by an overall suppression factor 
$\det S \sim |\nu \Box \picl|^{-2}\sim (r/r_V)^3$, arising from the Jacobian determinant.
\item
Any power of $k^2\Box\dpi$, resulting from the expansion of the exponential of (\ref{exponent})
is multiplied by the same power of $S^2$ after the rescaling of $k$ is performed. In the context of standard perturbation theory,
UV divergent terms $(\Box\dpi)^l$ 
would be enhanced by the presence of the background, as we discussed at the end of section \ref{perturb}. 
Within our scheme, they are suppressed by powers of $S$. 
\item
A possible enhancement is generated by the factor $\sim |\nu \Box \picl | \sim (r_V/r)^{3/2}\gg 1$ in eq. (\ref{exponent}). 
However, terms involving this factor also include powers of $\ex$ from the operators
$D_\ex$ and $L^{\mu\nu}_\ex$. As the power of $\ex$ cannot exceed 2 (for a logarithmic divergence),
the maximal enhancement is limited to a multiplicative factor
$\sim |\nu \Box \picl|^4 \sim (r_V/r)^6$. 
This is always overcompensated by the powers of $S$.
\item
A further enhancement can be possibly generated in the expansion of the exponential when derivative operators incorporated in 
$D_\ex$ and $L_\ex^{\mu\nu}$ act on $G_{\mu\nu}$.   
Again, this enhancement is limited by the requirement that $\ex$ do not exceed 2.
\end{itemize}

The general conclusion that can be reached by the above considerations is that the one-loop divergences of the theory are not 
enhanced by arbitrary powers of $\nu \Box \picl$ as the perturbative result (\ref{schem}) would imply. 
Moreover, suppression factors are generated through the modified calculation of the
heat kernel that we employed. They can be viewed as a result of the strong wavefunction renormalization induced by the
background. We demonstrated the strong suppression of the quartically divergent term, quadratic in the field.
In the following section we  consider the less divergent terms in a more general setting. 

\section{A general framework} \label{general}

We consider a fluctuation operator of the form
$\Delta = -G_{\mu\nu} \partial^\mu \partial^\nu$, 
with $G_{\mu\nu}$ a functional of the scalar field $\pi$ and its derivatives. For the operator to be self-adjoint, we require
$\partial^\mu G_{\mu\nu}=0$. Notice that this condition is satisfied by the operator (\ref{lapll}) when written in
the above form, with $ G_{\mu\nu}$ given by eq. (\ref{metric}).
Following the steps of section \ref{perturb},
the one-loop effective action (\ref{gamma1}) can be obtained from the heat kernel, whose diagonal part takes the form
\be
h(x,x,\ep)=\int\frac{d^4k}{(2 \pi)^4}\frac{1}{\ex^2}\exp \left\{
 - G_{\mu\nu}k^\mu k^\nu+2i\sqrt{\ex} G_{\mu \nu} k^\mu\partial^\nu+\ex G_{\mu \nu} \partial^\mu\partial^\nu
 \right\},
\label{diagongen}
\ee
with the implicit assumption that it acts on $f(x)=1$. The term
$\exp(-G_{\mu\nu}k^\mu k^\nu)$ can be isolated through use of the 
Baker-Campbell-Hausdorff formula (\ref{bch}). The terms up to order $\ex^2$ in the 
expansion of the remaining exponentials incorporate all possible UV divergences. The ``metric" $G_{\mu\nu}$ can be 
diagonalized as in eq. (\ref{diag}), while the momentum integrations can be
carried out through use of formulae such as (\ref{iden2}).  Finally, the effect of a nontrivial background can be
studied by writing the field as $\pi=\picl+\dpi$ and expanding in powers of $\dpi$.

It is straightforward to check that the quartically divergent terms of the cubic Galileon theory are generated correctly through this procedure. 
At order $\ex^{-2}$ the diagonal part of the heat kernel is simply
\be
h(x,x,\ex)=\frac{1}{16\pi^2}\frac{1}{\ex^2} \det S,
\label{heatgen} \ee
while from eq. (\ref{diag}) we obtain $\det S=(\det G)^{-1/2}$. In order to obtain the terms quadratic in $\dpi$ we need the 
expansion
\be
\det[G_0+\dG]=\det G_0 \left( 
1+{\rm tr}[G_0^{-1}\, \dG]-\frac{1}{2} {\rm tr}[(G_0^{-1}\, \dG)^2]+\frac{1}{2} \left({\rm tr}[G_0^{-1}\, \dG]\right)^2 + ...
\right).
\label{expdet} \ee
For the cubic Galileon theory
\begin{eqnarray}
(G_0)_{\mu \nu}&=&g_{\mu \nu}-2\nu\Box \pcl\, g_{\mu \nu}+2 \nu \partial_\mu \partial_\nu \pcl
\label{genmet0} \\
\dG_{\mu \nu}&=&-2\nu\Box \dpi\, g_{\mu \nu}+2 \nu \partial_\mu \partial_\nu \dpi.
\label{genmetd} 
\end{eqnarray}
For a spherically symmetric background, with $(G_0)_{\mu\nu}$ given by eq. (\ref{cart}), the effective action (\ref{eff2})
is reproduced. It is straighforward to extend the expansion in eq. (\ref{expdet}) in order to obtain terms of higher order
in $\dpi$. Such terms will be suppressed by additional powers of $\nu\Box \picl$. This should be contrasted with the
expectations from standard perturbation theory for the quartically divergent terms, as given by eq. (\ref{schem}) and discussed at
the end of section \ref{perturb}.

In order to compute the quadratically divergent terms, we employ the Baker-Campbell-Hausdorff formula (\ref{bch})
with
\be
X=-G_{\mu \nu} k^\mu k^{\nu},
\label{hi}
\ee
\be 
Y=2i\sqrt{\ep} G_{\mu \nu}  k^{\mu} \partial{^\nu}+\ep G_{\mu \nu} \partial^{\mu} \partial^{\nu}.
\label{psi}
\ee
The following commutators are needed in order to calculate the heat-kernel up to $\mathcal{O}( \ep)$: 
\be
[X,Y ]= 2i\sqrt{\ep} G_{\mu \nu} k^\alpha k^\beta k^\nu \partial^\mu G_{\alpha \beta}
+2\ep G_{\mu \nu} k^\alpha k^\beta \partial^\mu G_{\alpha \beta} \partial^{\nu}
+\ep G_{\mu \nu}  k^\alpha k^\beta \partial^{\mu} \partial^{\nu} G_{\alpha \beta}
\label{com1} \ee
\be
[ [X,Y], X ]  =-2 \ep   G_{\mu \nu}k^\alpha k^\beta k^\gamma k^\delta \partial^\mu G_{\alpha \beta } \partial^{\nu} G_{\gamma \delta}
\label{com2} \ee 
\be
[[X,Y],Y] =4 \ep G_{\kappa \lambda} k^\alpha k^\beta  k^{\lambda} k^{\nu} \partial^\kappa \left(G_{\mu \nu} \partial^{\mu} G_{\alpha \beta} \right)
+\mathcal{O} ( \ep^{3/2} )
\label{com3} \ee
\be
Y[X,Y]=-4 \ep G_{\kappa \lambda} k^\alpha k^\beta  k^{\lambda} k^{\nu} \partial^\kappa \left(G_{\mu \nu} \partial^{\mu} G_{\alpha \beta} \right)
-4\ex G_{\kappa \lambda} G_{\mu \nu} k^\alpha k^\beta k^\lambda k^\nu \partial^\mu G_{\alpha \beta} \partial^\kappa
+\mathcal{O} ( \ep^{3/2} )
\label{com4} \ee
\be
[X,Y][X,Y]=-4 \ep G_{\kappa \lambda} G_{\mu \nu} k^\alpha k^\beta k^\gamma k^\delta k^\lambda k^\nu \partial^\mu G_{\alpha \beta } \partial^{\kappa} G_{\gamma \delta}+\mathcal{O} (  \ep^{3/2} ).
\label{com5} \ee
It can be shown  that terms with more than two nested commutators, like $[[[X,Y],X],Y]$, are   $\mathcal{O}(\ex^{3/2})$.
If we expand the exponentials in eq. (\ref{bch}) up to the desired order we have
\be
e^{X+Y}=e^X \left(1-\frac{1}{2}Y[X,Y]-\frac{1}{2}[X,Y] +\frac{1}{8}[X,Y][X,Y] - \frac{1}{3} [[X,Y],Y]-\frac{1}{6}[[X,Y],X]+ \mathcal{O} ( \ex^{3/2} ) \right).
\ee
In order to perform the integration in eq. (\ref{diagongen}) we rescale the momenta as $k^\mu=S^\mu_{~\nu} k'^\nu$,
with $S^\mu_{~\nu}$ given by eq. (\ref{diag}), and employ formulae such as (\ref{iden2}). We also use the identity 
\bea
\int \frac{d^nk}{(\pi)^{n/2}} e^{- k^2}k^\alpha  k^\beta  k^\gamma k^\delta  k^{\lambda} k^\nu &=&
\frac{1}{8} \left( g^{\alpha \beta } g^{\gamma \delta} g^{\lambda \nu}+ 2g^{\alpha \beta } g^{\gamma \lambda} g^{\delta  \nu}
+2g^{\alpha \gamma } g^{\beta \delta} g^{\lambda \nu}  \right. \nn \\
&&+ \left. 4 g^{\alpha \gamma } g^{\beta \lambda} g^{\delta \nu}+4g^{\alpha \gamma } g^{\beta \nu} g^{\delta \lambda}+
2g^{\alpha \lambda } g^{\beta \nu} g^{\gamma \delta} \right),
\eea
where we have assumed that the $k$'s are to be contracted with a tensor that is symmetric with respect to the exchange of $\alpha$, $\beta$ and 
$\gamma$, $\delta$, respectively.

We finally obtain
\bea
h(x,x,\ep)&=&\frac{1}{(4 \pi \ep)^2}\det S \Biggl[  1+\frac{\ex}{48}
 \Bigl(
-4\Tc^{\alpha\beta}G_{\mu\nu}\, \partial^\mu\partial^\nu G_{\alpha\beta}       
+4\Tc^{\alpha\gamma} \partial^\mu G_{\alpha\kappa}\, \partial^\kappa G_{\gamma\mu} 
\Bigr. \Biggr. 
\nonumber \\
\Bigl. \Biggl.
 &&
+ G_{\mu\nu} \Tc^{\alpha\beta} \Tc^{\gamma\delta} \partial^\mu G_{\alpha\beta}\, \partial^\nu G_{\gamma\delta}
+2 G_{\mu\nu}\Tc^{\alpha\gamma} \Tc^{\beta\delta} \partial^\mu G_{\alpha\beta} \, \partial^\nu G_{\gamma \delta}
 \Bigr)          +\mathcal{O} ( \ex^{3/2} ) \Biggr],
\eea
with  ${\Tc}\equiv S^TS=G^{-1}$ and  $\det S=(\det G)^{-1/2}$. We have also made use of the condition  
$\partial^\mu G_{\mu\nu}=0$ in order to simplify the result.

We now concentrate on the form of the effective action around the 
spherically symmetric solution $\picl(r^2)$ given by eq. (\ref{solgsta}). The background ``metric" is given by eq. (\ref{cart}).
In the region $1/\Lx \lta r\lta r_V$, the elements of the ``metric" behave as
$G_{0} \sim (r/R_V)^{-3/2} $, while those of its inverse as $\Tc=S^T S \sim  (r/R_V)^{3/2}$. We also have $\det S \sim (r/R_V)^3$.
We write the field as $\pi = \pcl +\delta \pi$ and expand in powers of $\dpi$. We focus on the terms $\sim \dpi^2$.
Even though their explcit form is very complicated, it is possible to  deduce their general  structure through the following observations.

Every derivative operator in eq. (\ref{psi}) comes with one power of $\sqrt{\ex}$. 
This is the reason why there are two derivatives in every term of the heat kernel at ${\mathcal O}(\ex)$. 
The two derivative operators also come, schematically, with $G$ or $G^2kk$. The latter combination becomes $ G^2 \Tc$ 
after the integration of the momenta. Contributions from the $X$ of eq. (\ref{hi}) always produce factors of 
$G\Tc$ after the integration of the momenta.
The general conclusion is 
that  the power of $G$ minus the power of $\Tc$ is one in every term. As a result, 
every term will have the same power law behavior for $r\lta R_V$ when evaluated on the background. 
Another important observation is that every derivative operator produces a factor of $r^{-1}$ when acting on the background. 

Let us now consider the splitting $\pi = \pi_{cl}+\delta \pi$.  
The factors of $\dpi$ result from the expansion of $G$, $\Tc$ and $\det S$. 
In all cases the appearance of one power of $\dpi$ effectively reduces the power of $G$ minus the power of $\Tc$ by one. 
This is obvious for factors of $G$. For factors of $\Tc$ it results from the relation
\be
\delta \Tc^{\mu \rho}=-\Tc^{\sigma \rho} \Tc^{\mu\nu} \delta G_{\mu\nu},
\label{texp} \ee
while for $\det S$ it results from eq. (\ref{expdet}). It must also be noted that every factor of $\dpi$ comes with two derivatives acting 
on it, as can be seen from eq. (\ref{genmetd}). 
It may also have an additional derivative if it orginates in a term of the type $\partial G$, or it may have two additional derivatives 
if it orginates in a term $\partial^2 G$. In these last two cases, one or two factors of $r^{-1}$ are lost relative to the counting of the
previous paragraph, which assumed that all derivatives act on the background.
The upshot of this reasoning is that the general form of the effective action is 
\bea
\Gamma^{(2)}_{1}&=&\nu^2\int d^4x \left( c_0\frac{r^6}{R_V^6} \Lambda^4 \left( \delta \pi \partial^4 \delta \pi \right) 
+c_{1a} \frac{r^{5/2}}{R_V^{9/2}} \Lambda^2 \left( \delta \pi \partial^4 \delta \pi \right) +\right. \nn \\  
&&+ \left. c_{1b} 
 \frac{r^{7/2}}{R_V^{9/2}} \Lambda^2\left(\delta \pi \partial^5 \delta \pi \right)
+ c_{1c} \frac{r^{9/2}}{R_V^{9/2}} \Lambda^2\left(  \delta \pi  \partial^6 \delta \pi \right)
+\mathcal{O}\left(\log\Lx \right)\right),
\eea
with $c_i$ dimensionless constants. As we have already emphasized, the above expression is schematic. It indicates only the number of 
derivatives acting on $\dpi$ in each term and the effect of the background. The exact index
structure of the derivative operators has been omitted. For comparison, the exact form of the quartically divergent term is given
by eq. (\ref{heat4}).

Using the same arguments it is possible to deduce the general form of the logarithmically divergent term of the heat kernel, obtained at
order $\ex^2$ in the expansion of the exponential (\ref{bch}). 
In this case there are four-derivative operators, accompanied by $G^2$ or $G^3 \Tc$ or $G^4 \Tc^2$. 
The power of $G$ minus the power of $\Tc$ is now two. The four derivatives contribute a factor of $r^{-4}$ when they act on the
background. 
Splitting the field as $\pi = \pi_{cl}+\delta \pi$ and following the logic of the previous paragraph, we find 
\bea
\Gamma^{(2)}_{\ex^2} &=& \nu^2\int d^4x \log  (\Lambda/\mu) 
\biggl(
c_{2a}  \frac{1}{rR_V^3}   \left(\delta \pi \partial^4 \delta \pi \right)
+c_{2b}\frac{1}{R_V^3}   \left(  \delta \pi \partial^5 \delta \pi \right)
+c_{2c}  \frac{r}{R_V^3}  \left(  \delta \pi  \partial^6 \delta \pi \right) 
\biggr.
\nonumber \\
&&
\biggl. 
+c_{2d}\frac{r^2}{R_V^3}   \left(  \delta \pi \partial^7 \delta \pi \right)
+c_{2e}  \frac{r^3}{R_V^3}  \left(  \delta \pi  \partial^8 \delta \pi \right) 
\biggr).
\label{logg}\eea

It is apparent that the terms that involve higher powers of $\dpi$ will be
more suppressed by the background than the quadratic terms. The reason is
that each power of $\dpi$ results from the expansion of 
$G$, $\Tc$ and $\det S$, with the simultaneous
reduction of the power of $G$ minus the power of $\Tc$ by one. 
As factors of $G$ enhance the action by powers of $(R_V/r)^{3/2}$,
their elimination suppresses the result. This behavior is completely 
opposite to what would be expected from the perturbative result (\ref{schem}),
as we discussed at the end of section \ref{perturb}.

It must be pointed out that the calculation of $\trr \log \Delta$ for a fluctuation operator of the form 
$\Delta = -G_{\mu\nu} \partial^\mu \partial^\nu$ can be mapped to the corresponding calculation for a similar operator with 
{\it covariant} derivatives involving both a Riemann and a gauge part. The explicit correspondence is provided in section 2 of ref. \cite{vassilevich}.
In our case, the gravitational and gauge backgrounds would become complicated functions of the scalar field $\pi$ and its derivatives.  
This approach has the advantage that immediate use can be made of known results for the one-loop effective action \cite{salcedo}.
However, translating these results into expressions for the effective action of the fluctuation $\dpi$ around a Galileon background $\picl$ 
is rather technical because of the complicated correspondence between the two pictures. 
A similar approach has been followed in refs. \cite{sigma} in order to compute 
the one-loop corrections in the nonlinear sigma model. The nonlinear realization of the symmetry can be
 exploited in order to
consider the fields as coordinates on a coset manifold that possesses a set of Killing vectors. 
The one-loop effective action can be constructed from the Killing vectors in a way that exhibits its geometric properties.
Calculations along such  lines 
would provide an independent cross-check of our conclusions and will be the focus of future work.

\section{Conclusions} \label{discussion}

Our analysis leads to the remarkable conclusion that the one-loop quantum corrections can be suppressed in 
certain regions of an inhomogeneous background. This is a feature not encountered in renormalizable theories. For example, one 
may consider a domain-wall background in a renormalizable scalar theory with a double-well potential. The background will 
influence the quantum corrections through the effective mass term of the fluctuations $m(\picl)$. In order to repeat our calculation,
we would redefine the momenta in the heat kernel as $k'^2=k^2+m^2(\picl)$. This change of integration variable would have
no significant effect on the UV divergences, as long as $\Lx \gg m(\picl)$.

The question of whether the quantum effects may be suppressed on the background of classical configurations has also been
addressed in the context of classicalization \cite{dvali,rizos}. 
This proposal concerns
the nature of high-energy scattering in certain classes of nonrenormalizable scalar field theories. It advocates that
scattering can take place at length scales much larger than the
typical scale associated with the nonrenormalizable terms in the Lagrangian. Quantum corrections are expected to be subleading
at such scales, so that a semiclassical description should be sufficient. The inspiration is taken from ultra-Planckian
scattering in gravitational theories, during which a black hole is expected to start forming at distances comparable to the
Schwarzschild radius. The analogue of the black hole is a semiclassical configuration, the classicalon, generated by a point-like source.
In ref. \cite{vikman} it was argued that quantum fluctuations in $\dpi$ can be suppressed 
for theories, such as the ``wrong-sign" DBI theory, that admit classicalons.

The application of our approach to classicalizing theories is a direction for future research.
A general class of models that can support classicalons has an action
of the form   
\be
S=\int d^4x \, {\cal K}  \left( X \right),
\label{genaction}
\ee
with $X=\partial_{\mu} \pi \partial^{\mu} \pi /2$. 
The DBI action corresponds to  $ {\cal K}=\mu^{-1} \sqrt{1+2\mu X}$.
The
second variation of the action (\ref{genaction}) results in the fluctuation
operator
\be 
\Delta=-\calk_{X} \, \Box- \calk_{XX} \, \partial_{\mu} \pi \partial{_\nu} \pi \,  \partial^{\mu} \partial ^{\nu}
-\left(\partial_{\mu} \calk_{X}+\partial^{\nu} \calk_{XX} \partial_{\mu} \pi\partial _{\nu}\pi+\calk_{XX} \partial_\mu \partial^{\nu} \pi
\partial_{\nu} \pi + \calk_{XX} \partial_{\mu} \pi \Box \pi\right)\, \partial^{\mu} ,
\ee
where $\calk_{X}= \calk'\left(X\right), \, \calk_{XX}=\calk'' \left(X\right)$.
The analysis of the one-loop quantum corrections through the heat 
kernel around a classicalon configuration can be performed along the lines we followed in this work. 
In this way a deeper understanding of the phenomenon of classicalization can be obtained.

We conclude by noting that both classicalization and the Vainshtein mechanism rely on strong nonlinear effects 
associated with the background. The combined picture arising through our work and ref. \cite{vikman}
supports the speculation that the suppression of quantum 
effects by the background may be a usual phenomenon in higher-derivative theories.

\section*{Acknowledgments}
We would like to thank S. Sibiryakov and A. Vikman for useful discussions.
The work of N.T. has been supported in part 
by the European Commission under the ERC Advanced Grant BSMOXFORD 228169.
The research of N.B. and N.T. has been co-financed by the European Union (European Social Fund – ESF) and Greek national 
funds through the Operational Program ``Education and Lifelong Learning" of the National Strategic Reference 
Framework (NSRF) - Research Funding Program: ``THALIS. Investing in the society of knowledge through the 
European Social Fund".

\end{document}